\documentclass{article}
\usepackage{spconf,amsmath,graphicx,amssymb,multirow}

\title{Multi-Cycle-Consistent Adversarial Networks for CT Image Denoising}
\name{
\begin{tabular}{@{}c@{}}
Jinglan Liu$^1$\thanks{This research was approved by the Research Ethics Committee of Guangdong General Hospital, Guangdong Academy of Medical Science with the protocol No. 20140316.}, Yukun Ding$^1$, Jinjun Xiong$^{2}$, Qianjun Jia$^{3}$,\\ Meiping Huang$^3$, Jian Zhuang$^3$,Bike Xie$^4$, Chun-Chen Liu$^4$, Yiyu Shi$^1$
\end{tabular}
}

\address{$^1$ Department of Computer Science and Engineering, University of Notre Dame, USA\\
$^2$ IBM Thomas J. Watson Research Center, USA\\
$^3$ Guangdong General Hospital, China\\
$^4$ Kneron Inc., USA}
\begin{document}
%
\maketitle
\begin{abstract}
CT image denoising can be treated as an image-to-image 
translation task where the goal is to learn 
the transform 
between a source domain $X$  (noisy images) and 
a target domain $Y$ (clean images). Recently, 
cycle-consistent adversarial denoising network (CCADN) 
has achieved state-of-the-art results by enforcing 
cycle-consistent loss without the need 
of paired training data. 
Our detailed analysis of CCADN raises a number of 
interesting questions. 
For example, if the noise is large leading 
to significant 
difference between domain $X$ and domain $Y$, 
can we bridge $X$ and $Y$ with an intermediate 
domain $Z$ such that both the denoising process 
between $X$ and $Z$ and that between $Z$ and $Y$ are 
easier to learn? As such intermediate domains 
lead to multiple cycles, how do we best enforce 
cycle-consistency? Driven by these questions, 
we propose a multi-cycle-consistent adversarial 
network (MCCAN) that builds intermediate domains and 
enforces both local and global cycle-consistency. 
The global cycle-consistency couples all generators 
together to model the whole denoising process, 
while the local cycle-consistency imposes effective 
supervision on the process between adjacent domains. 
Experiments show that both local and global 
cycle-consistency are important for the 
success of MCCAN, which outperforms the state-of-the-art.
\end{abstract}
\section{Introduction}
\label{sec:intro}

Computed tomography (CT) is one of the most widely used 
medical imaging modality for showing anatomical 
structures 
\cite{you2018structurally,liu2019machine,xu2019accurate,xu2019whole}. 
The foremost 
concern of CT examination is the associated exposure 
to radiation, which is known to increase the lifetime 
risk for death of cancer \cite{hobbs2018physician}. 
The radiation dose can be lowered at the cost 
of image quality 
\cite{you2018structurally}, 
and the resulted 
images are denoised for enhanced perceptual quality 
and diagnostic confidence from radiologists. 
 
Various deep neural network (DNN) based methods
exist for CT image 
denoising \cite{shan20183,wolterink2017generative,chen2017low,yang2018low}, which require paired clean 
and noisy images for training. 
Yet simulations are usually used to generate 
such paired data, where the synthetic noise patterns 
can be different from the real ones, leading to 
biased training results \cite{kang2018cycle}. 
To address this issue, recently 
cycle-consistent adversarial denoising 
network (CCADN) was  
proposed in \cite{kang2018cycle}, which 
formulates CT image denoising as an 
image-to-image translation 
problem without paired training
data.
CCADN consists of two generators: 
one transforms noisy CT images (domain $X$) 
to clear ones (domain $Y$) and 
the other transforms clear CT images (domain $Y$) 
to noisy ones (domain $X$). 
Both generators are trained 
by adversarial 
loss. In addition, 
cycle-consistency loss 
and identity loss are utilized 
to gain better performance 
\cite{zhu2017unpaired}, which 
will be discussed in detail in 
Section~\ref{method}. However, 
since CCADN only 
contains two domains $X$ and $Y$, 
its efficacy 
degrades as the noise becomes 
stronger leading to larger
differences between $X$ and $Y$ 
that are harder to learn.

To tackle this issue, we propose to 
establish an intermediate domain 
between the original 
noisy image domain $X$ and clear image domain 
$Y$, 
and decompose the denoising task into multiple 
coupled steps such that each step is easier 
to learn by DNN-based models.
Specifically, we construct an additional domain $Z$ 
with images of intermediate noise level between 
$X$ and $Y$. These images can be 
considered as a step stone in the 
denoising process and provide additional 
information for the training of the 
denoising network. 
The multi-step framework particularly suits the 
denoising problem: while it is difficult to 
either find or define a good collection of 
images in the ``half-cat, half dog" domain 
in ``cat-to-dog'' type of image translation 
problems, a domain $Z$ of images with 
intermediate 
level of noise exist naturally.

With the new domain $Z$, we further propose a 
multi-cycle-consistent adversarial network to 
perform the multi-step denoising, which 
builds multiple cycles of different 
scales (global cycles and local cycles) 
between the domains while 
enforcing the corresponding cycle-consistencies. 
In the experiments, we find that both 
global cycles and local cycles are 
necessary for the success of MCCAN, 
which combined outperforms the state-of-the-art 
competitor CCADN.

\section{Methodology}
\label{method}
Given training images that are either labelled as noisy 
(domain $X$) or clear (domain $Y$), we first construct 
a new domain $Z$ which contains images with an 
intermediate noise level between $X$ 
and $Y$. How to obtain $Z$ is flexible in practice. 
In our experiments, it is obtained from 
$X$ and $Y$ by 
separating out those images 
with intermediate noise 
level. 

\begin{figure}[htb]

\begin{minipage}[b]{.48\linewidth}
  \centering
  \centerline{\includegraphics[height=0.65in]{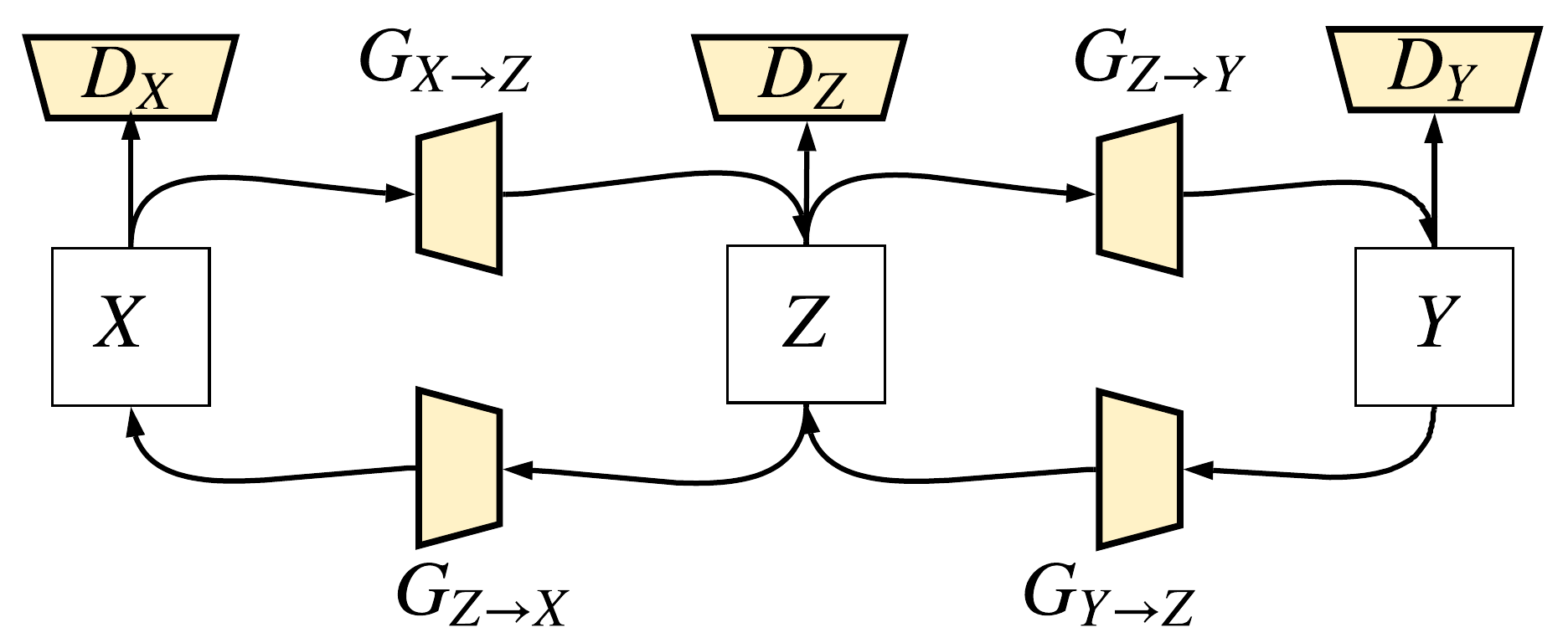}}
  \centerline{(a)}\medskip
\end{minipage}
\hfill
\begin{minipage}[b]{0.48\linewidth}
  \centering
  \centerline{\includegraphics[height=0.63in]{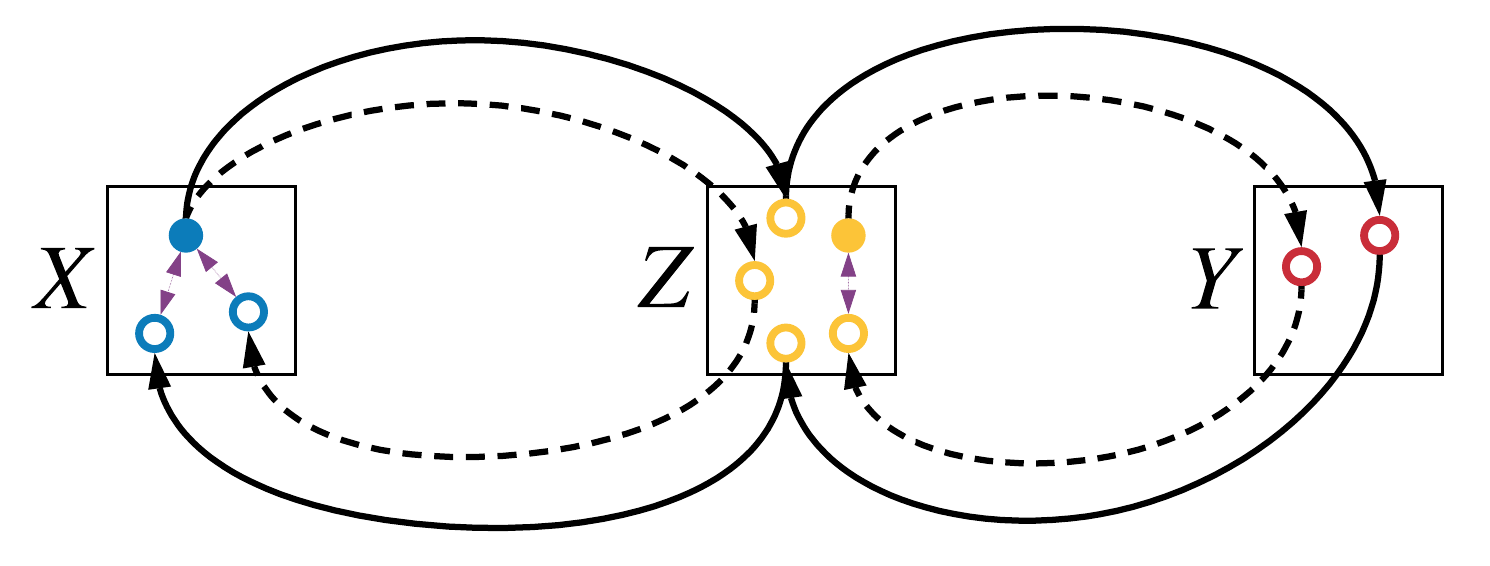}}
  \centerline{(b)}\medskip
\end{minipage}
\caption{(a) Structure of MCCAN and (b) its cycles. The arrows inside each domain denote the computation of cycle-consistency loss. The solid and dashed arrows across domains form global and local cycles, respectively. For clarity, we only show cycles from left to right. Symmetric cycles going from right to left also exist but are not shown.}
\label{fig:comp}

\end{figure}

With CT images from three domains, 
the multi-step 
denoising architecture of MCCAN 
is shown in Fig.~\ref{fig:comp}(a). We train four 
convolutional neural networks as generators and 
three as discriminators. 
Arrows in Fig.~\ref{fig:comp}(a) define how images 
are transformed in the training stage. Specifically, 
the generator $G_{X\to Z}$ aims 
to transform an image from $X$ to $Z$. 
$G_{Z\to X}$, $G_{Z\to Y}$, and 
$G_{Y\to Z}$ can be 
interpreted similarly. Discriminators 
$D_X$, $D_Y$, 
and $D_Z$ aim to distinguish the ``real'' 
images originally belonging to the domains 
$X$, $Y$, and $Z$ 
from the ``fake'' images transformed from 
other domains respectively. 

As the MCCAN structure in 
Fig.~\ref{fig:comp}(a) contains 
thee domains, there are multiple 
ways in which we can construct 
cycles (paths where an image from a 
source domain is transformed through 
one (in \cite{zhu2017unpaired}) 
or several other domains (in this paper) 
and back to the source domain) for 
cycle-consistent loss. 
In particular, we introduce two types of 
cycles as shown in Fig.~\ref{fig:comp}(b). 
In this figure, each dot represents an 
image, which is color-coded based on the domain. 
The solid ones represent the images 
originally in the domain 
(``real'' ones), and the hollow ones 
represent those transformed 
from another domain (``fake'' ones).  
As such, the dashed arrows form the 
{\bf local cycles}, each of which 
goes across only two adjacent domains. 
On the other hand, the solid arrows 
constitute a {\bf global cycle} that 
starts from $X$ through 
$Z$, $Y$, $Z$, and back to $X$ sequentially. 
Note that in the figure we only show 
half of the cycles (from left to right) 
for clarity, and the other half which 
are from right to left and symmetric 
to the ones shown also exist. 
We then enforce cycle-consistency 
loss, which measures 
the difference between the original 
images and the final images 
produced at the end of the 
cycle as represented by 
the small arrows within 
each domain in Fig~\ref{fig:comp}(b). 
Ideally, the images transformed 
back to the source domain should 
be identical to the original ones. 
The cycle-consistency loss is 
applied to every cycle, no matter 
whether it is local or a global.

The global 
cycles are important for the denoising 
performance due to the following reason. 
In the inference stage, an input noisy 
CT image $x$ in domain $X$ will be transformed 
by $G_{X\to Z}$ and $G_{Z\to Y}$ sequentially, 
which means $G_{X\to Z}$ and $G_{Z\to Y}$ 
are coupled by data dependency. Without 
global cycles, $G_{X\to Z}$ and $G_{Z\to Y}$ 
will be trained independently. Thus, errors of 
the prediction of noise at intermediate steps 
may be accumulated as processing progresses. 
The global cycles enable the joint training 
of the generators, which models the denoising 
path used in the inference stage for 
better consistency. 

The local 
cycles are also important to address two issues in the training. 
First, the global cycles go through 
all the four generators and 
have long paths for the gradient 
to back-propagate, which makes the 
end-to-end optimization 
difficult. The locals cycles are shallow and have 
shorter paths for the gradient to 
back-propagate. Second, 
adversarial training only enforces 
the generators to output 
``fake'' images identically distributed as 
the original ``real'' images in the 
intermediate domain $Z$. 
However, they do not necessarily preserve 
the meaningful content in the inputs, 
which is critical for the denoising task. 
The local cycle-consistency supervises each generator 
to learn to transform images while preserving their 
meaningful content from the inputs more easily.

In summary, our MCCAN has two major advantages 
over CCADN. First, it decomposes the one-step transform 
into multiple steps using images in a constructed 
intermediate domain as a step stone. Second, it not only 
incorporates global cycles that model the denoising path in 
the inference stage for consistency, but also uses local 
cycles that provide strong supervision to facilitate the 
more challenging training process. In the experiments we 
find that MCCAN outperforms CCADN.


Note that in the discussion so far, only one 
intermediate domain was assumed. It is also 
possible to include more than one intermediate 
domains with more global and local cycles. However, 
our study suggests that any additional domains beyond 
one will not introduce further performance 
gain in the dataset we explored.

Finally, we state the training objective used in our framework. 
Denote $\{G\}$ and $\{D\}$ as the set of generators 
and discriminators respectively. 
Denote $I\in \{X,Y,Z\}$ as one domain and $D_I$ as the 
discriminator associated with domain $I$. 
We let $C_i$ be a cycle and $P_{i,j}$ be a path 
of half $C_i$ that has the same source domain, 
where $i, j$ are used to distinguish different 
cycles and paths merely. 
For example, $X\rightarrow Z\rightarrow X$ is 
a cycle, saying $C_1$, 
thus we can have $P_{1,1} = X \rightarrow Z$, 
and $P_{1,2} = Z \rightarrow X$, which are both 
half cycles of $C_1$. 
$\{P_{I}\}$ represents the 
set of all the paths that end at 
domain $I$. We denote $I_{C_i}$ as the 
source domain of $C_i$ and 
$G_{P_{i,j}}$ as the ordered function composition 
of the generators in $P_{i,j}$. Thus, 
the total adversarial loss is

\begin{align}
\label{equ:loss_gan_xy}
    \mathcal{L}_{GAN}(\{G\}, \{D\}) =\sum_{ I\in\{X,Y,Z\}} \sum_{ P_{i,j}\in\{P_I \}}\mathcal{L}_{GAN}(I,P_{i,j})
\end{align}
where $\mathcal{L}_{GAN}(I,P_{i,j})$ is the adversarial loss associated with domain $I$ and the transform path $P_{i,j}$.  $\mathcal{L}_{GAN}(I,P_{i,j})$ is obtained by
\begin{align}
\label{equ:loss_gan_xy_i}
\begin{split}
    \mathcal{L}_{GAN}(I,P_{i,j}) &=
    \mathbb{E}_{y\sim p_{data}(I)}
    [\log{D_{I}(y)}]\\
    &+ \mathbb{E}_{x\sim p_{data}(I_{C_i})}
    [log(1-D_{I}(G_{P_{i,j}}(x)))]
\end{split}
\end{align}
where $p_{data}(I)$ is the distribution of 
``real'' images in the domain $I$ and $D_I(x)$ 
represents the probability determined by 
$D_I$ that $x$ is a ``real'' image from 
domain $I$ rather than a ``fake'' one 
transformed by generators from another 
domain. 

The cycle-consistency loss is associated 
with each $C_i$, defined as
\begin{align}
\mathcal{L}_{\text{cyc}}(\{G\},C_i) =  \mathbb{E}_{x\sim p_{\text{data}}(I_{C_i})}[|G_{C_i}(x)-x|_1].
\end{align}

The final optimization problem we solve in 
the training stage is:
\begin{equation}
\begin{split}
    \{G\}^* &= \arg\min_{\{G\}}\max_{\{D\}}( \mathcal{L}_{GAN}(\{G\}, \{D\})\\
    &+\lambda \sum_{C_i\in \{C\}} \mathcal{L}_{cyc}(\{G\}, C_i).
\end{split}
\end{equation}
where $\lambda$ is set to 10 in our experiments. 


\section{Experiments and Results}
\subsection{Experiments Setup}
\label{DataGeneration}

The original dataset contains 200 
normal-dose 3D CT images 
and 200 low-dose ones from 
various patients for training, and separate 
11 images for test. All examinations 
are performed with a wide detector
256-slice MDCT scanner (Brilliance iCT; 
Philips Healthcare) providing 8cm of 
coverage. 
Each 2D CT image is of size  
512$\times$512, which is then
randomly cropped into 256$\times$256 
for data augmentation. 
We construct the additional domain $Z$ with 
images of intermediate noise
level from these clear and noisy scans 
to make the number of scans in each 
domain comparable. There are CT images with 
more noise than usual from clear scans that 
use high dose radiation, and vice versa, 
because the noise variation cannot be 
controlled quantitatively. 

\begin{figure}[htb]
\centering
\begin{minipage}[b]{0.48\linewidth}
\centerline{\includegraphics[height=0.5in]{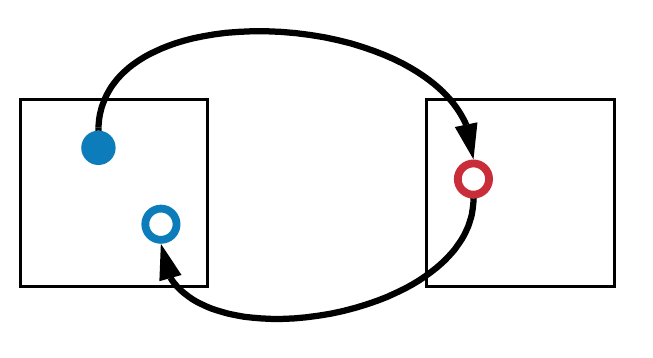}}
\centerline{(a)}\medskip
\end{minipage}
\hfill
\begin{minipage}[b]{0.48\linewidth}
\centerline{\includegraphics[height=0.5in]{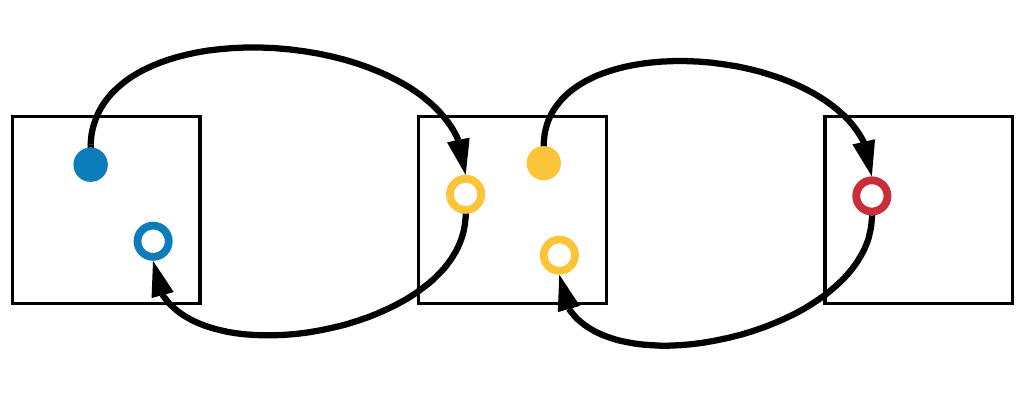}}
\centerline{(b)}\medskip
\end{minipage}

\begin{minipage}[b]{0.48\linewidth}
\centerline{\includegraphics[height=0.5in]{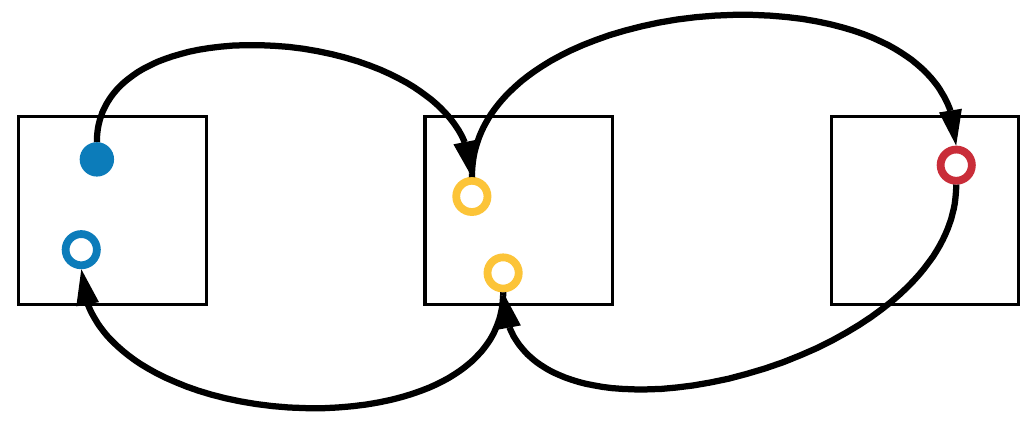}}
\centerline{(c)}\medskip
\end{minipage}
\begin{minipage}[b]{0.48\linewidth}
\centerline{\includegraphics[height=0.5in]{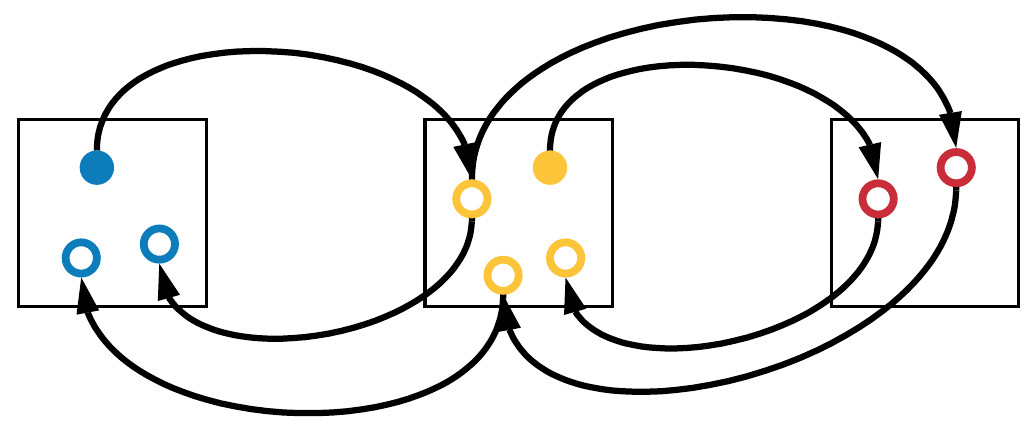}}
\centerline{(d)}\medskip
\end{minipage}
\vspace{-.2in}
\caption{Comparison of (a) CCADN, (b) MCCAN without global cycles (c) MCCAN without local cycles, and (d) MCCAN. For the clarity of presentation, we only show cycles from left to right and symmetric cycles from right to left also exist.}
\label{fig:compexp}
\end{figure}

We compare MCCAN with a state-of-the-art 
CT denoising framework 
CCADN \cite{kang2018cycle}. 
In order to see how the local cycles and 
global cycles contribute to the final 
performance, we also implement and compare 
MCCAN without local cycles 
and without global cycles 
respectively as ablation study. 
The various structures are 
shown in Fig. \ref{fig:compexp}. 
We train all the networks 
following the setting in 
\cite{zhu2017unpaired}. Our 
implementation will be available 
online. 
We ensure that all 
network sizes and number of 
training epochs are the same 
for fair comparisons.

\begin{figure}
\vspace{-.02in}
\begin{minipage}[b]{1.\linewidth}
  \centering
  \centerline{\includegraphics[height=0.9in]{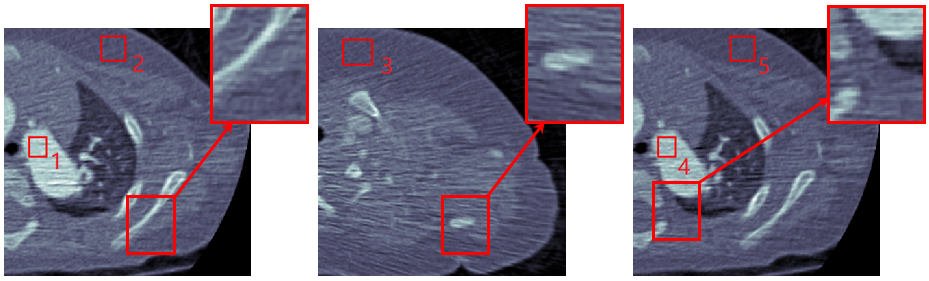}}
  \centerline{(a) Original noisy CT images}\medskip
\end{minipage}

\begin{minipage}[b]{1.\linewidth}
  \centering
  \centerline{\includegraphics[height=0.9in]{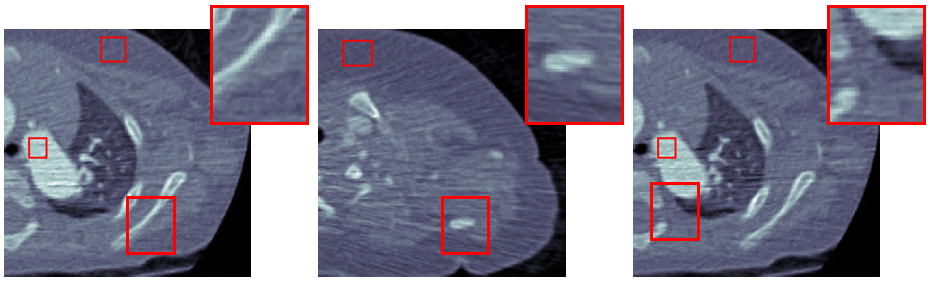}}
  \centerline{(b) Images denoised by CCADN\cite{kang2018cycle}}\medskip
\end{minipage}
\vspace{-.02in}
\begin{minipage}[b]{1.\linewidth}
  \centering
  \centerline{\includegraphics[height=0.9in]{images/miccai/cg.PNG}}
  \centerline{(c) Images denoised by MCCAN without global cycles}\medskip
\end{minipage}
\vspace{-.02in}
\begin{minipage}[b]{1.\linewidth}
  \centering
  \centerline{\includegraphics[height=0.9in]{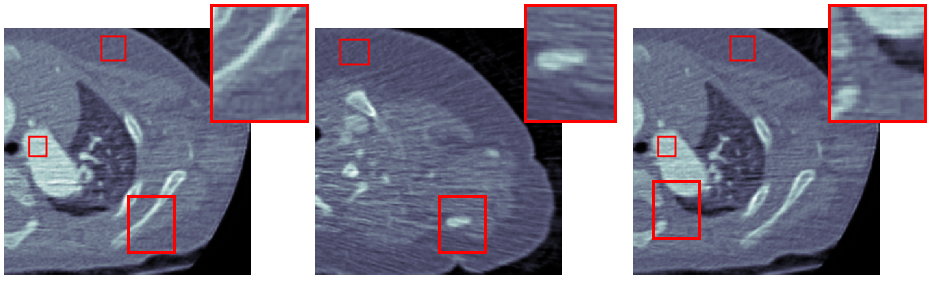}}
  \centerline{(d) Images denoised by MCCAN without global cycles}\medskip
\end{minipage}
\vspace{-.02in}
\begin{minipage}[b]{1.\linewidth}
  \centering
  \centerline{\includegraphics[height=0.9in]{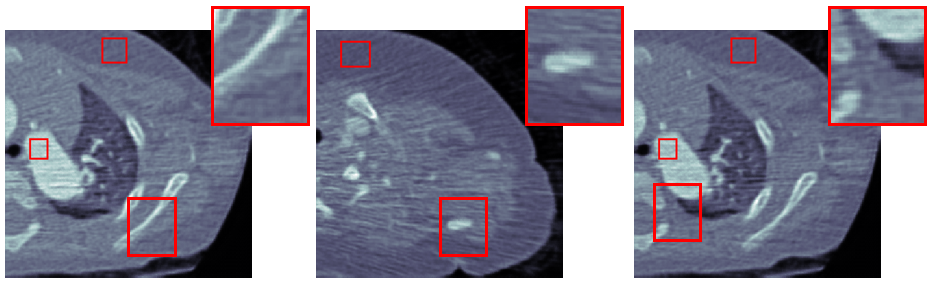}}
  \centerline{(e) Images denoised by MCCAN}\medskip
\end{minipage}
\vskip -0.2in
\caption{Noisy and denoised 
images for qualitative evaluation.}
\label{fig:overview}
\end{figure}
\begin{figure}
\centering
\includegraphics[width=3.2in]{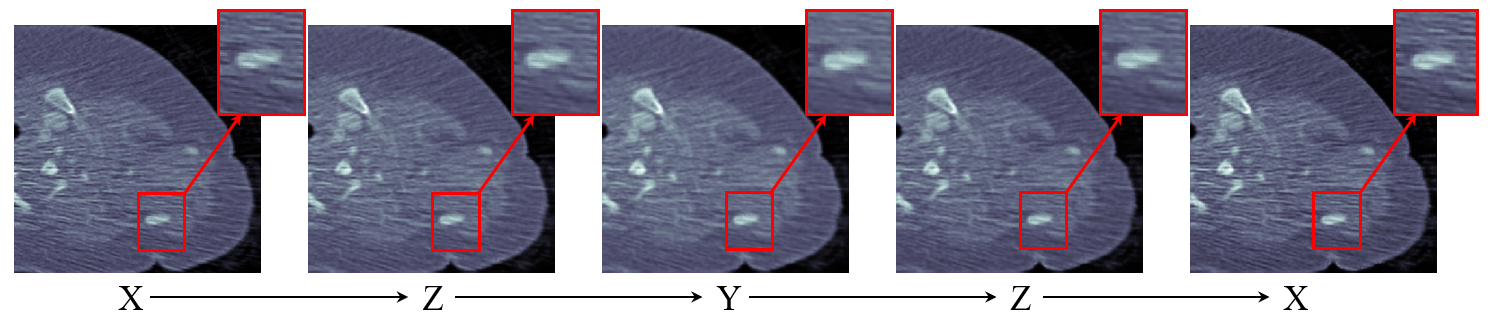}
\vspace{-.2in}
\caption{An image transformed through X$\rightarrow$Z$\rightarrow$Y$\rightarrow$Z$\rightarrow$X cycle in Fig.~\ref{method}. The noise level decreases along X$\rightarrow$Z$\rightarrow$Y and increases along Y$\rightarrow$Z$\rightarrow$X, which conforms to our design.}
\label{fig:ImageInCycle}
\end{figure}
\subsection{Qualitative Evaluation}
We choose three representative low-dose 
CT images in the test dataset as shown in 
Fig.~\ref{fig:overview}(a) for qualitative evaluation. 
The corresponding denoised images by 
CCADN, MCCAN without local cycles, 
MCCAN without global cycles, and MCCAN are shown in Fig.~\ref{fig:overview}(b)-
\ref{fig:overview}(e) 
respectively. 
Numbered areas are 
homogeneous regions, while areas 
with edges between heterogeneous 
regions are zoomed for visibility 
in Fig. \ref{fig:overview}. 
From the figures we can see that
CCADN can successfully reduce noise in the 
original images. MCCAN without local 
cycles completely fails to produce 
reasonable results. A more closer 
examination of the images reveal that 
interestingly the background and the 
substances are approximately swapped 
compared with the original images. This is 
because the high-level 
features of content distribution are 
still kept even with such swap,
and the discriminator cannot identify 
the generated image as 
``fake'' because of the structure diversity in 
the training dataset. This aligns with 
our discussion on the importance of 
local cycles in Section~\ref{method}.   
On the other hand, MCCAN without global cycles 
can successfully 
denoise the image and achieves similar 
quality compared with CCADN. 
This is expected as MCCAN without global 
cycles is essentially formed 
by two cascaded CCADNs. 
Finally, with both local and global 
cycles, the complete MCCAN 
has the smallest noise visually. 

\begin{table*}[htb]
  \begin{center}
  \caption{Mean and SD (normalized) of the selected areas in Fig.~\ref{fig:overview}(a). }
  \label{table:199_idose}
  \begin{tabular}{c|c|c|c|c|c|c|c|c|c|c}
  \hline
  \multirow{2}{*}{Method} & \multicolumn{2}{c|}{Area \#1} & \multicolumn{2}{c|}{Area \#2} & \multicolumn{2}{c|}{Area \#3} & \multicolumn{2}{c|}{Area \#4} & \multicolumn{2}{c}{Area \#5}\\
  \cline{2-11}
  & Mean & SD & Mean & SD & Mean & SD & Mean & SD & Mean & SD\\
  \hline
  Original & 1.00 & 1.00 & 1.00 & 1.00 & 1.00 & 1.00 & 1.00 & 1.00 & 1.00 & 1.00\\
  \hline
  CCADN\cite{zhu2017unpaired} & 0.93 & 0.85 & 1.03 & 0.79 & 1.03 & 0.80 & 0.94 & 0.78 & 1.03 & 0.78\\
  \hline
  MCCAN w/o local cycles & 0.02 & 0.38 & 0.24 & 1.02 & 0.11 & 0.71 & 0.02 & 0.42 & 0.28 & 1.31\\
  \hline
  MCCAN w/o global cycles & 0.89 & 0.78 & 1.03 & 0.77 & 1.03 & 0.80 & 0.906 & 0.73 & 1.03 & 0.81\\
  \hline
  MCCAN & 0.88 & \textbf{0.76} & 1.04 & \textbf{0.68} & 1.03 & \textbf{0.71} & 0.89 & \textbf{0.71} & 1.04 & \textbf{0.68}\\
  \hline
  \end{tabular}
  \end{center}
\vskip -0.3in
\end{table*}

To further illustrate the efficacy of the 
MCCAN structure, Fig.~\ref{fig:ImageInCycle} 
shows how an image is transformed along a 
global cycle (the path
X$\rightarrow$Z$\rightarrow$Y$\rightarrow$Z$\rightarrow$X). From the figure we can 
see that $X\rightarrow Z \rightarrow Y$ is 
an effective two-step denoising process 
while $Y \rightarrow Z \rightarrow X$ 
incrementally adds noise back.  

\subsection{Quantitative Evaluation}
Following existing works \cite{wolterink2017generative,yang2018low,arapakis2014using}, 
we use the mean and standard deviation (SD)
of pixels in homogeneous regions of
interest chosen by radiologists 
to quantitatively judge the quality 
of CT images.
The mean value reflects substance information.
Although the closer to that in the origin 
image the better, mean value can fluctuate 
within a range.
On the other hand, the standard 
deviation reflects the noise level. 
It should be as low as possible, which is 
more sensitive than the mean value 
in the denoising task. 

Five homogeneous areas   
chosen by radiologist are used
for the quantitative evaluation, which 
are annotated by red rectangles in 
Fig.~\ref{fig:overview} and numbered from 1 to 5. 
The normalized quantitative results are shown in 
Table \ref{table:199_idose}. 
CCADN can reduce the standard deviation in the five 
areas by 15\%, 21\%, 
21\%, 22\% , and 22\% 
respectively, with resulting 
mean values close to those of the original images.
Although MCCAN without local cycles 
achieves smallest standard 
deviation in Areas 1, 3 and 4, it leads to 
meaningless output with large mean deviation 
from the original images, which corresponds 
to the structure loss 
in Fig.~\ref{fig:overview}(c).
MCCAN without global cycles has 
similar performance 
compared with CCADN. 
with mean values close to original and standard 
deviation reduction by 22\%, 23\%, 20\%, 
27\%, and 19\% respectively. 
Finally, the complete MCCAN behaves the 
best among all the methods: 
Within reasonable mean range, 
the standard deviations 
are decreased the most 
by 24\%, 32\%, 29\%, 29\%, and 32\% 
from the original CT images respectively.

\section{Conclusions}
In this paper, we propose multi-cycle-consistent
adversarial network (MCCAN) 
for CT image denoising. MCCAN builds 
intermediate domains and enforces both 
local and global cycle-consistency. 
The global cycle-consistency 
couples all generators together to model the 
whole denoising process, while 
the local cycle-consistency imposes 
effective supervision on the denoising process 
between adjacent domains. 
Experiments show that both local and global 
cycle-consistency are important for the 
success of MCCAN and it outperforms 
the state-of-the-art competitor.




\bibliographystyle{IEEEbib}
\bibliography{strings,refs}

\end{document}